%% file: pmu_fm_paper.tex
\documentclass[conference]{IEEEtran}
\IEEEoverridecommandlockouts
\usepackage{cite}
\usepackage{amsmath,amssymb}
\usepackage{graphicx}
\usepackage{booktabs,array}
\usepackage{balance}
\usepackage{tikz}
\usetikzlibrary{arrows.meta,positioning,fit,calc}
\usepackage[hidelinks]{hyperref}
\renewcommand{\IEEEauthorrefmark}[1]{%
  \raisebox{0pt}[0pt][0pt]{\textsuperscript{\footnotesize #1}}}

\begin{document}
\bstctlcite{IEEE:BSTcontrol}

\title{PMU Foundation Model Bridge: A Tool for Connecting
	Synchrophasor Streams to Time Series Foundation Models%
	\thanks{\copyright~2027 IEEE. Personal use of this material is permitted.
		Permission from IEEE must be obtained for all other uses, in any current
		or future media, including reprinting/republishing this material for
		advertising or promotional purposes, creating new collective works,
		for resale or redistribution to servers or lists, or reuse of any
		copyrighted component of this work in other works.}}

\author{\IEEEauthorblockN{Haozong Wang\IEEEauthorrefmark{1},
Jiahui Yang\IEEEauthorrefmark{1},
Yuru Wu\IEEEauthorrefmark{1,2},
Chujie Zeng\IEEEauthorrefmark{1},
Qian Liu\IEEEauthorrefmark{1},\\
Yan Wen\IEEEauthorrefmark{1},
Yayu Yang\IEEEauthorrefmark{1},
Muhammad Umar Afzaal\IEEEauthorrefmark{1}, and
Yilu Liu\IEEEauthorrefmark{1,2}}
\IEEEauthorblockA{\IEEEauthorrefmark{1}The University of Tennessee, Knoxville, TN, USA}
\IEEEauthorblockA{\IEEEauthorrefmark{2}Oak Ridge National Laboratory, Oak Ridge, TN, USA}}

\maketitle

\begin{abstract}
This paper presents PMU Foundation Model Bridge, an open source tool that connects online phasor measurement unit (PMU) or phasor data concentrator (PDC) streams to publicly available time series foundation models (TSFM). Through a desktop interface or command line interface, users can select measurement channels, define preprocessing, choose TSFMs, and configure outputs. In the initial release, 4 publicly available TSFMs were supported, including Chronos-2, TimesFM 2.5, Timestral, and TimesFM-3. The tool supports both file and synchrophasor streaming outputs, therefore enabling both offline use and integration into C37.118-based online workflow applications. In addition to supporting raw synchrophasor data, the framework provides a preprocessing layer for deriving customized additional quantities for forecasting. The tool is validated using a virtual PMU and real-time laboratory measurements from a frequency disturbance recorder (FDR). The tool aims to support community research on combining synchrophasor technology with TSFMs.
\end{abstract}

\begin{IEEEkeywords}
Synchrophasor, phasor measurement unit, time series foundation model, online forecasting.
\end{IEEEkeywords}

\section{Introduction}
Forecasting is an important task in power system planning and operation. Examples include load and wind power forecasting for energy scheduling~\cite{wang2025} and electricity price forecasting for market decisions~\cite{zhang2022}. For applications concerned with faster grid dynamics, synchrophasor technology provides an important source of synchronized voltage, current, and frequency measurements through phasor measurement units (PMUs)~\cite{you2017non}. These measurements support prediction of future system states when communication delays leave some measurements unavailable~\cite{yu2019}, as well as prediction of system dynamics following disturbances~\cite{pei2023}.

Traditional forecasting approaches often describe temporal dependencies using predefined signal or statistical models, such as autoregressive models~\cite{pei2023}. In the deep learning era, dramatic increases in model complexity and parameter amounts have enabled superior performance on complex forecasting tasks~\cite{yu2019,deepar}. However, these methods are generally developed for a particular task, with model selection, parameter fitting, and training tailored to that application. In recent years, large language models (LLMs) based on the Transformer architecture have demonstrated the potential of pretraining on large datasets to support multiple tasks with limited adaptation to each task~\cite{chang2024survey}. Motivated by these advances, time series foundation models (TSFMs) based on Transformers learn reusable temporal patterns from diverse collections of time series data, treating forecasting across domains as a shared learning problem. These models can therefore forecast new series without training for each specific task.

Connecting synchrophasor streams with TSFMs is therefore a natural option for exploring power system forecasting applications. However, practical integration involves several issues in online data handling and model execution, including the formation of historical windows from measurements that may arrive out of order, incomplete or invalid data, and variable inference latency. Further considerations include the temporal interpretation of forecast outputs and their channel mappings for downstream applications, as well as maintaining responsive operation during slow inference or interrupted input.

To support the power systems research community, this paper presents PMU Foundation Model Bridge, an open source tool that connects online PMU or phasor data concentrator (PDC) streams to Chronos-2~\cite{chronos2}, TimesFM 2.5~\cite{timesfm25}, TimesFM-3~\cite{timesfm3}, or Timestral~\cite{timestral}. It provides channel selection, optional Python preprocessing, local inference, and network streaming or file output through a desktop interface and a command line interface (CLI). The source code and usage instructions are publicly available on GitHub.\footnote{\url{https://github.com/wanghaozong88/pmu-fm}} The paper describes the tool's architecture and functional verification with virtual PMU and laboratory measurement streams. The tool aims to support research on integrating synchrophasor technology with TSFMs.

\section{Model Inputs, Outputs, and Timing}
\subsection{Foundation Model Inputs and Outputs}
The foundation model takes a recent history of selected signals and predicts their future values. For forecast $r$, the input matrix $\mathbf{X}_r\in\mathbb{R}^{N\times C}$ has $N$ rows for historical sampling times and $C$ columns for channels. The nominal sample interval after preprocessing is $\Delta$. The pretrained model $F_\theta$ predicts the next $H$ samples of each channel:
\begin{equation}
\widehat{\mathbf{Y}}_r=F_\theta(\mathbf{X}_r;H),
\qquad \widehat{\mathbf{Y}}_r\in\mathbb{R}^{H\times C}.
\label{eq:forecast}
\end{equation}
The output contains $H$ predicted samples for each of the $C$ channels. Model parameters $\theta$ remain fixed during use.

Users can select channels and set the history length, forecast length, and interval between model calls.

\subsection{Key Times and Their Relationships}
Each predicted value is associated with two key timestamps: a reference time marking the end of the input and a target time indicating when the prediction applies. The reference time $t_r$ is the measurement time of its last input sample. The $k$th predicted sample has target time
\begin{equation}
t_{r,k}^{\mathrm{target}}=t_r+k\Delta,\qquad k=1,\ldots,H.
\label{eq:target}
\end{equation}
where the difference $k\Delta$ is the forecast lead.

\begin{figure}[t]
\centering
\input{figures/timing.tex}
\caption{Time relationships for one forecast.}
\label{fig:timing}
\end{figure}
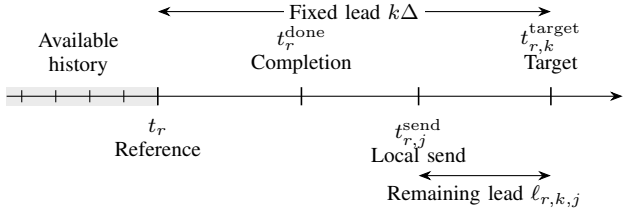

Receiving data, waiting, computing, and sending reduce the time remaining before the target. Denote $t_r^{\mathrm{done}}$ as the timestamp when computation finishes and $t_{r,j}^{\mathrm{send}}$ as the timestamp when local transmission to client $j$ finishes. Fig.~\ref{fig:timing} shows these four times. At local send completion, the remaining lead is

\begin{equation}
\ell_{r,k,j}=t_{r,k}^{\mathrm{target}}-t_{r,j}^{\mathrm{send}}.
\label{eq:lead}
\end{equation}

For a meaningful forecast, $\ell_{r,k,j}>0$ is a necessary condition, assuming aligned source and local clocks. However, it should be noted that local send completion does not confirm downstream receipt. Therefore, a positive remaining lead is not a sufficient condition for timely delivery.

\section{System Architecture}
\begin{figure*}[t]
\centering
\resizebox{\textwidth}{!}{\input{figures/architecture.tex}}
\caption{Tool architecture. Solid arrows show data flow, and dashed arrows show configuration and supervision. Preprocessing, the selected model, and an optional inverse transform run in one worker process. File and network outputs use the same forecast record.}
\label{fig:architecture}
\end{figure*}
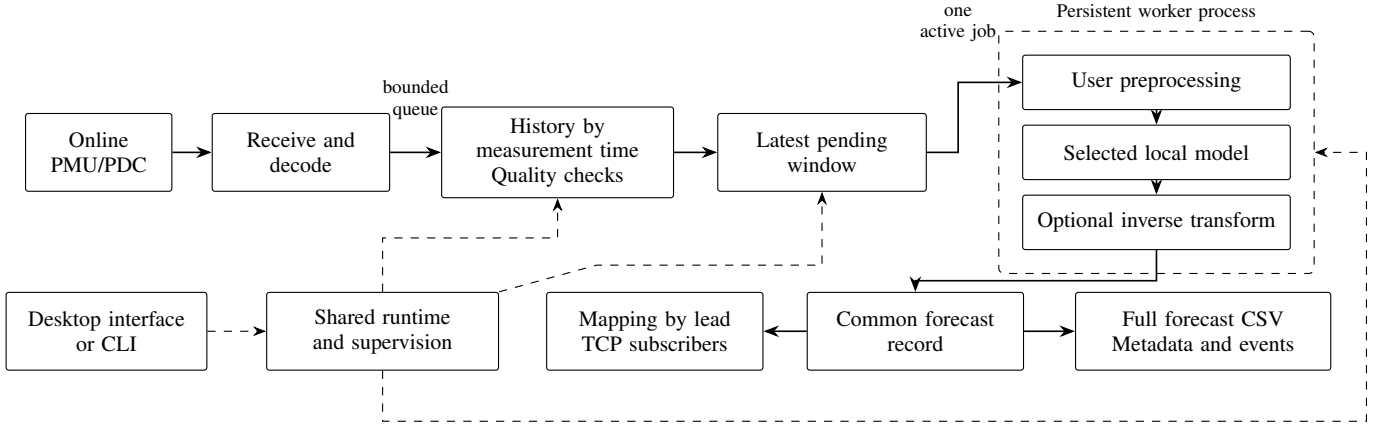

Fig.~\ref{fig:architecture} shows how data move through the tool: incoming measurements form historical windows, preprocessing prepares the model inputs, and completed forecasts are sent to file and network outputs. A common runtime coordinates these steps. The desktop interface and CLI both configure this runtime and display its status.

Before a run, users can select among Chronos-2, TimesFM 2.5, Timestral, or TimesFM-3 and provide its local weight directory. The tool loads the selected model once for successive forecasts. A model adapter handles the required input format and returns predictions in a common format. All four backends forecast each channel independently. Covariate support and joint multivariate forecasting can be added in future work.

\subsection{Historical Windows}
The receiver reads the source configuration to identify the available channels and their units. It stores the selected values together with measurement times, reception times, and quality information. Samples are sorted by measurement time in a buffer with a fixed capacity. A short waiting period allows for samples arriving out of order, and each submitted window keeps a fixed set of samples. Duplicate and excessively late samples are discarded and counted.

Before prediction, the tool checks the window's sampling intervals, completeness, and quality. By default, windows containing missing, nonfinite, invalid, or unsynchronized data are skipped, with the reason recorded. Users can enable limited repair of missing data through preprocessing. For sources with rounded timestamps, samples are matched to expected times counted backward from the window reference. A small configured tolerance allows for rounding, with unique matches required. The tolerance is bounded relative to the source and model intervals. Accepted samples retain their original timestamps and values.

\subsection{Preprocessing}
Despite raw synchrophasor data, users may also wish to forecast customized quantities calculated from PMU measurements, such as variables or indicators defined for a particular study. The preprocessing layer lets users construct these time series and pass them to the foundation model. This extends the tool to forecasting tasks involving both reported measurements and quantities defined by users.

An optional Python function receives a copy of the selected historical data and descriptions of their times, units, sources, and quality. Users can calculate new channels through these data. The function returns the processed time series along with the corresponding channel information, which could be selected as inputs of the TSFMs. The tool checks the returned times, values, and channel mappings before prediction, and carries the names and units into the results. If preprocessing normalizes the input, users can provide an inverse function to restore predictions to the original scale and units. 

\subsection{Handling Delays and Interruptions}
The worker processes one forecast at a time. When inference takes longer than the interval between model calls, new eligible windows can become available while the worker is still busy. The runtime keeps one window waiting for execution. If a newer eligible window becomes available before the waiting window is submitted, it replaces the waiting window, and the replacement is recorded. This gives priority to recent measurements.

The runtime monitors execution time and worker memory. A monotonic clock measures elapsed time for timeout checks. If execution exceeds a time or memory limit, the worker is terminated. Restart attempts are limited. If an input interruption leads to a reconnection, the tool will clear the history and collect a fresh window. A change in source configuration stops the run for channel remapping. Results from a previous source state are discarded.

Input reception, file writing, and network sending use separate queues with fixed capacities. These activities can proceed independently. A client is disconnected if its queue fills or sending exceeds the time limit, and file errors or dropped records are reported. When the tool's run ends, it closes its input and output connections, stops the worker, and closes the output files. Each downstream client can use START and STOP commands to enable or pause forecast transmission on its own connection.

\subsection{Output Structure}
The tool supports both file and network outputs. It contains the predicted values, channel descriptions, reference and target times, completion time, and model information. Each forecast has a unique identifier, allowing successive forecasts for the same target time to be stored separately. All four backends return median point forecasts, with this statistic recorded alongside the values.

For file outputs, CSV files will be generated to store the complete forecast sequence. Accompanying metadata identify channel sources and units, the model and weights used, and the preprocessing script. Publication events record submission for sending and completion of local sends. File writing status is tracked separately from network transmission.

For network outputs, the tool generates a C37.118 synchrophasor stream and places selected forecast leads in ANALOG channels in floating point format. Each channel represents a reported or calculated quantity at a fixed lead, and the frame timestamp is $t_r$. A mapping record links each output channel to its name, unit, source, and forecast lead. Downstream applications use this mapping to interpret the predictions.

The FREQ and DFREQ fields carry source measurements at the same reference time. Frames are published when these required source values are valid, with a status flag indicating that the data have been modified. Skipped publications are recorded as events.

\section{Validation}
\subsection{Virtual PMU Stream}
\begin{figure}[t]
\centering
\includegraphics[width=\columnwidth]{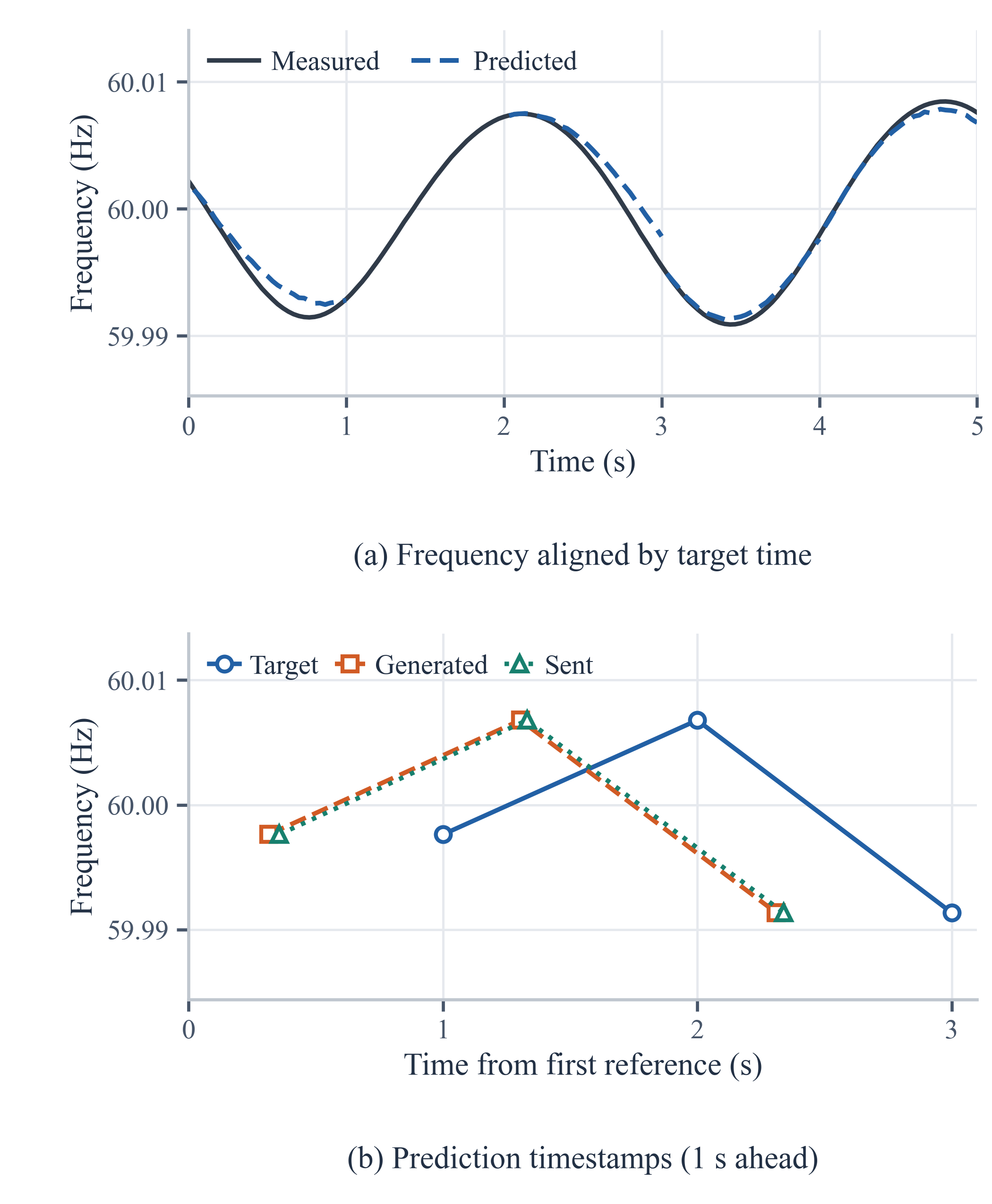}
\caption{Basic validation using TimesFM 2.5: (a) measured and predicted frequency and (b) target, completion, and local send times of predictions with a lead of one second.}
\label{fig:frequency-validation}
\end{figure}

An initial test was conducted using a virtual PMU. The virtual PMU ran on the same workstation as the tool and supplied an online measurement stream, which was processed using TimesFM 2.5 to generate frequency forecasts. The resulting forecast stream was transmitted to PMU Connection Tester, where the reception of configuration and forecast data frames was verified using its parser. This test covered the processing chain from measurement acquisition and local forecasting to downstream reception.

Fig.~\ref{fig:frequency-validation}(a) compares the recorded frequency with predictions aligned by target time. The predictions follow the main oscillation, with differences in parts of the record. Fig.~\ref{fig:frequency-validation}(b) shows three predictions with a lead of one second. Each predicted value is plotted at its target $t_{r,k}^{\mathrm{target}}$, computation completion $t_r^{\mathrm{done}}$, and local send completion $t_{r,j}^{\mathrm{send}}$. Computation and local transmission both finished before the target times for the three forecasts shown.

\subsection{Laboratory PMU Stream}
An additional test used live measurements received from a laboratory frequency disturbance recorder (FDR). A configuration frame was imported to decode the online stream. The measurements arrived at 10 samples/s. TimesFM 2.5 used 4~s of history to predict the next second, with one model call per second. In the test, the tool predicted frequency, rate of change of frequency, voltage magnitude, and voltage angle.

\begin{figure}[t]
\centering
\includegraphics[width=\columnwidth]{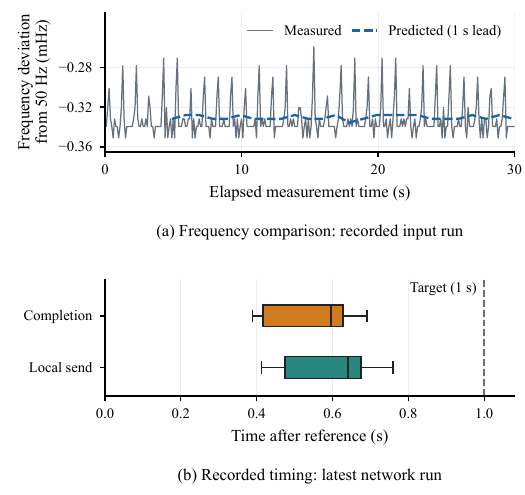}
\caption{Laboratory FDR tests using TimesFM 2.5: (a) measured frequency and predictions from a run with saved input windows; (b) completion and local send times relative to the reference in a later network run. Boxes span interquartile ranges, with median lines and full-range whiskers. The vertical line marks the 1 s target. The panels use separate runs on the same day.}
\label{fig:laboratory-fdr}
\end{figure}

Fig.~\ref{fig:laboratory-fdr}(a) compares measured frequency with predictions. Predictions with a lead of one second are matched to measurements at the same target times. Fig.~\ref{fig:laboratory-fdr}(b) summarizes the timing of the test. Computation finished 0.39--0.69~s after the reference times, with a median of 0.60~s. Local TCP transmission finished 0.41--0.76~s after the reference times, with a median of 0.64~s. All local sends completed before their corresponding targets with a lead of one second, leaving 0.24--0.59~s until the target. These results show that the tool processed the live stream and completed local transmission within the selected forecast lead during this run.

\balance
\section{Conclusion}
PMU Foundation Model Bridge provides a configurable connection between online synchrophasor streams and publicly available time series foundation models. The tool combines channel selection, flexible preprocessing, local model execution, and file and network outputs in a shared runtime. Users can work with reported or calculated quantities and choose among four supported models through the same workflow. Tests with a virtual PMU and laboratory PMU data demonstrate online model execution and output generation. The tool is intended to support community research on integrating synchrophasors with TSFMs. Future development can add covariates and joint multivariate forecasting and extend validation to a wider range of power system data.

\section*{Acknowledgment}
The authors acknowledge that the software code for PMU Foundation Model Bridge was developed by OpenAI Codex.

\bibliographystyle{IEEEtran}
\bibliography{references}
\end{document}

%% file: figures/timing.tex
\begin{tikzpicture}[x=1cm,y=1cm,font=\footnotesize,>=Stealth]
\fill[black!8] (0,-0.12) rectangle (2.0,0.12);
\draw[->] (0,0) -- (8.15,0);
\foreach \x in {0.2,0.65,1.1,1.55} {\draw (\x,-0.08)--(\x,0.08);}
\foreach \x in {2.0,3.9,5.45,7.2} {\draw[semithick] (\x,-0.12)--(\x,0.12);}
\node[align=center] at (0.95,0.56) {Available\\history};
\node[align=center,anchor=north] at (2.0,-0.18) {$t_r$\\Reference};
\node[align=center,anchor=south] at (3.9,0.18) {$t_r^{\mathrm{done}}$\\Completion};
\node[align=center,anchor=north] at (5.45,-0.18) {$t_{r,j}^{\mathrm{send}}$\\Local send};
\node[align=center,anchor=south] at (7.2,0.18) {$t_{r,k}^{\mathrm{target}}$\\Target};
\draw[<->] (2.0,1.12) -- node[fill=white,inner sep=2pt] {Fixed lead $k\Delta$} (7.2,1.12);
\draw[<->] (5.45,-1.05) -- node[below] {Remaining lead $\ell_{r,k,j}$} (7.2,-1.05);
\end{tikzpicture}

%% file: figures/architecture.tex
\begin{tikzpicture}[
  font=\footnotesize,
  box/.style={draw, rounded corners=1pt, align=center, minimum height=10mm, inner sep=4pt},
  flow/.style={-{Stealth[length=2mm]}, semithick},
  control/.style={-{Stealth[length=1.7mm]}, dashed},
  note/.style={font=\scriptsize, align=center}
]
\node[box, text width=16mm] (source) at (0.9,0) {Online\\PMU/PDC};
\node[box, text width=20mm] (receive) at (3.5,0) {Receive and\\decode};
\node[box, text width=27mm] (history) at (6.8,0) {History by\\measurement time\\Quality checks};
\node[box, text width=24mm] (pending) at (10.2,0) {Latest pending\\window};
\node[box, text width=33mm, minimum height=7mm, inner sep=2pt] (preprocess) at (14.5,0.9) {User preprocessing};
\node[box, text width=33mm, minimum height=7mm, inner sep=2pt] (model) at (14.5,0) {Selected local model};
\node[box, text width=33mm, minimum height=7mm, inner sep=2pt] (inverse) at (14.5,-0.9) {Optional inverse transform};
\node[draw, dashed, rounded corners=2pt, fit=(preprocess)(model)(inverse), inner sep=3mm,
  label={[note]above:Persistent worker process}] (boundary) {};
\node[box, text width=25mm] (record) at (11.4,-2.3) {Common forecast\\record};
\node[box, text width=25mm] (network) at (8.05,-2.3) {Mapping by lead\\TCP subscribers};
\node[box, text width=30mm] (files) at (15.1,-2.3) {Full forecast CSV\\Metadata and events};
\node[box, text width=27mm] (runtime) at (4.55,-2.3) {Shared runtime\\and supervision};
\node[box, text width=23mm] (ui) at (1.0,-2.3) {Desktop interface\\or CLI};
\draw[flow] (source) -- (receive);
\draw[flow] (receive) -- node[note,above=3mm] {bounded\\queue} (history);
\draw[flow] (history) -- (pending);
\draw[flow] (pending.east) -- (11.95,0) |- (preprocess.west);
\node[note,anchor=south] at (11.95,1.3) {one\\active job};
\draw[flow] (preprocess) -- (model);
\draw[flow] (model) -- (inverse);
\draw[flow] (inverse.south) -- (14.5,-1.65) -| (record.north);
\draw[flow] (record) -- (network);
\draw[flow] (record) -- (files);
\draw[control] (ui) -- (runtime);
\draw[control] (runtime.north) -- (4.55,-1.1) -| (history.south);
\draw[control] (runtime.north east) -- (7.3,-1.45) -- (10.2,-1.45) -- (pending.south);
\draw[control] (runtime.south) -- ++(0,-0.65) -| (17.2,0) -- (boundary.east);
\end{tikzpicture}